\documentclass[aps,prd,reprint,nofootinbib,superscriptaddress,floatfix]{revtex4-2}

\usepackage[T1]{fontenc}
\usepackage[utf8]{inputenc}
\usepackage{amsmath,amssymb,bm}
\usepackage{mathrsfs}
\usepackage{graphicx}
\usepackage{slashed}
\usepackage{xcolor}
\usepackage{hyperref}
\hypersetup{colorlinks=true,linkcolor=blue,citecolor=blue,urlcolor=blue}

\newif\ifdraftnotes
\draftnotesfalse

\begin{document}

\title{Gauge-invariant effective theory from the generalized uncertainty principle}

\author{Hector Gisbert}
\email{hector.gisbert@universidadeuropea.es}
\affiliation{Escuela de Ciencias, Ingenier\'ia y Dise\~no,
Universidad Europea de Valencia, Paseo de la Alameda 7,
46010, Valencia, Spain}

\author{Victor Ilisie}
\email{victor.ilisie@universidadeuropea.es}
\affiliation{Escuela de Ciencias, Ingenier\'ia y Dise\~no,
Universidad Europea de Valencia, Paseo de la Alameda 7,
46010, Valencia, Spain}

\author{Ezequiel Valero}
\email{ezequiel.valero@universidadeuropea.es}
\affiliation{Escuela de Ciencias, Ingenier\'ia y Dise\~no,
Universidad Europea de Valencia, Paseo de la Alameda 7,
46010, Valencia, Spain}
\affiliation{Facultat de F\'isica, Universitat de Val\`encia,
Carrer del Dr. Moliner, 50, 46100 Burjassot, Valencia, Spain}

\date{October 2026}

\begin{abstract}
We construct a local gauge-invariant effective theory for a
deformation inspired by the generalized uncertainty principle
(GUP), using a straight parallel transporter.
Within this prescription, the real three-ordering family satisfying
free-field matching and Hermiticity yields a common reduced action
at first order in the deformation, with correlated magnetic dipoles,
four-fermion vector-current interactions, and a CP-even cubic gauge interaction.
We also identify an off-shell gauge-invariance obstruction in a
previously proposed Abelian local action.
Tree-level matching gives $\Lambda_\beta\gtrsim5.1$ TeV from the
muon magnetic moment, conditional on the adopted Standard Model
prediction and a central Gaussian 95\% interval, and
$\Lambda_\beta\gtrsim3$ TeV from the published 95\% LEP limit
with unchanged weak interactions.
ATLAS and CMS dijet fits yield template-equivalent scales of
$7.1$ and $5.7$ TeV, respectively, rather than exclusions of
the correlated theory.
\end{abstract}

\maketitle

\section{Introduction}
\label{sec:introduction}

A generalized uncertainty principle (GUP) can introduce a minimum
position uncertainty through a deformation of the canonical
commutator~\cite{Maggiore:1993rv,Kempf:1994su}. Lorentz-covariant extensions have also been
developed~\cite{Quesne2006,Todorinov2019}.
Its implications for particle physics depend on how the
deformation enters the interactions. In higher-derivative
realizations, the free operator provides a matching condition.
Covariant derivatives, however, do not commute. Different
orderings therefore agree for free fields but differ in the
presence of gauge fields. Whether this freedom survives in
physical interactions is the question addressed in this work.

Gauge-theory realizations of GUP-inspired deformations have
been studied in
Refs.~\cite{Kober:2010sj,Kober:2011dn,Bosso:2020fos,Bosso2021QED}.
Covariance of the deformed derivative and gauge invariance
of a local action are separate requirements. The commutator
curvature contains derivatives acting to the right, so its
internal trace alone does not define a local scalar.
In Appendix~\ref{app:kober-test}, we show that the Abelian
local action displayed in Eq.~(22) of
Ref.~\cite{Kober:2010sj}, interpreted as written, fails the
off-shell gauge identity. The underlying differential
operator remains gauge covariant.

The radiative structure of a related GUP-deformed
QED realization has been investigated at one loop
in the massless limit~\cite{Valero2026QEDRen}.
After eliminating redundant operators, that
analysis identifies an independent axial--axial
four-fermion interaction required by renormalization.
Here, in contrast, we focus on the tree-level
operator correlations generated by a specified
local gauge-action prescription.

We consider the real three-ordering family in a vectorlike
gauge theory. Free-field matching and Hermiticity require
$a+b+c=1$ and $a=c$, leaving one coefficient arbitrary.
We construct the local gauge action by applying the complete
curvature product to a straight parallel transporter and
taking the coincidence limit. Parallel transporters are
standard tools in covariant derivative
expansions~\cite{Kuzenko2003,FuentesMartin2025Functional};
here their endpoint transformation law ensures gauge
invariance of the prescribed local action.

At first order in the deformation parameter $\beta$, the
ordering dependence cancels between the matter and gauge
sectors. We demonstrate this cancellation through a local
field redefinition~\cite{Criado2019Redefinitions}.
The reduced theory contains magnetic dipole interactions
proportional to the fermion masses, four-fermion vector-current interactions, and a CP-even cubic gauge interaction. For the
chosen prescription, their coefficients depend on a single
parameter, $\beta$. We check the dipole coefficient directly
in the electromagnetic vertex, including the external-state
normalization.

High-energy Compton scattering has been used to constrain
GUP-inspired effective interactions~\cite{Valero2025GUP}.
Here we use the correlated coefficients to interpret constraints
from lepton magnetic moments~\cite{Aliberti2025WP},
LEP contact interactions~\cite{Schael2013LEP}, and dijet
production~\cite{Alioli2017QCD}. The electromagnetic results
give bounds within a benchmark with unchanged weak
interactions. The dijet fits instead indicate sensitivity
to the four-quark interaction: their quadratic contributions
require information beyond the first-order expansion
considered here.

The paper is organized as follows.
Section~\ref{sec:framework} introduces the covariant
deformation and conventions. We reduce the fermion action
in Sec.~\ref{sec:fermion-sector} and construct the local
gauge action in Sec.~\ref{sec:gauge-sector}.
Section~\ref{sec:reduced-eft} presents the combined matching
and its ordering independence.
We discuss experimental constraints in
Sec.~\ref{sec:phenomenology} and conclude in
Sec.~\ref{sec:conclusions}.
The appendices provide the operator reductions,
the electromagnetic vertex calculation, and the
general gauge-invariance analysis.

\section{Theoretical framework}
\label{sec:framework}

\subsection{GUP-inspired deformation}
\label{subsec:free-deformation}

Deformations of the canonical position--momentum commutator
can introduce a minimum position uncertainty~\cite{Kempf:1994su}. A standard quadratic parametrization
in one spatial dimension, for states with vanishing mean momentum,
$\langle p\rangle=0$, is
\begin{equation}
\Delta x\,\Delta p
\geq
\frac{\hbar}{2}
\left[1+\beta_{\mathrm{GUP}}(\Delta p)^2\right].
\label{eq:gup-uncertainty}
\end{equation}
For $\beta_{\mathrm{GUP}}>0$, minimizing the lower bound on
$\Delta x$ gives $\Delta x_{\min}=\hbar\,\sqrt{\beta_{\mathrm{GUP}}}$.
This relation motivates a short-distance scale, while its implementation
in a relativistic field theory requires an explicit operator realization.

Working henceforth in natural units, $\hbar=c=1$, we adopt the
leading higher-derivative realization considered in
Ref.~\cite{Kober:2010sj},
\begin{equation}
\partial_\mu\longrightarrow\partial'_\mu
=\left(1-\beta\,\Box\right)\partial_\mu
+\mathcal O(\beta^2),
\qquad
\Box\equiv\partial^\nu\partial_\nu.
\label{eq:free-deformation}
\end{equation}
We use Eq.~\eqref{eq:free-deformation} as the free-field matching
condition and to fix the normalization of $\beta$. Relating $\beta$ to $\beta_{\mathrm{GUP}}$ requires an explicit
realization of the relativistic position and momentum operators.
We leave this relation unspecified.

For $\beta>0$, we define the suppression scale $\Lambda_\beta$ by
\begin{equation}
\beta\equiv\frac{1}{\Lambda_\beta^2}.
\label{eq:beta-scale}
\end{equation}
We treat the deformation perturbatively at momenta and masses
small compared with $\Lambda_\beta$.

\subsection{Gauge theory setup}
\label{subsec:conventions-assumptions}

We consider a vectorlike $SU(N)$ gauge theory with a Dirac
fermion $\psi$ of mass $m>0$ in the fundamental representation.
We work in Minkowski spacetime with
$\eta_{\mu\nu}=\operatorname{diag}(+,-,-,-)$ and
$\epsilon^{0123}=+1$. The Hermitian gauge potential is
$A_\mu=A_\mu^A T^A$, with
\begin{equation}
\operatorname{tr}_g(T^A T^B)=\frac12\delta^{AB},
\qquad [T^A,T^B]=if^{ABC}T^C.
\label{eq:gauge-generators}
\end{equation}
For the Abelian theory we use canonical Maxwell normalization.

Our covariant derivative and field strength are
\begin{align}
D_\mu&=\partial_\mu-igA_\mu,
\qquad [D_\mu,D_\nu]=-igF_{\mu\nu},
\nonumber\\
F_{\mu\nu}&=\partial_\mu A_\nu-\partial_\nu A_\mu
-ig[A_\mu,A_\nu].
\label{eq:gauge-conventions}
\end{align}
For a local gauge-matrix coefficient $M$, we define
\begin{equation}
\nabla_\rho M=\partial_\rho M-ig[A_\rho,M],
\quad D_\rho\circ M=M D_\rho+\nabla_\rho M.
\label{eq:operator-composition}
\end{equation}
Here $D_\rho$ acts on everything to its right, whereas
$\nabla_\rho M$ is a local coefficient. We also define
\begin{equation}
J_\mu\equiv\nabla^\nu F_{\mu\nu},
\qquad
\widetilde F^{\mu\nu}
=\frac12\epsilon^{\mu\nu\rho\sigma}F_{\rho\sigma},
\label{eq:field-strength-divergence}
\end{equation}
with $\nabla_\mu\widetilde F^{\mu\nu}=0$ by the Bianchi identity.
For spinors, we use $\bar\psi=\psi^\dagger\gamma^0$,
$\slashed D=\gamma^\mu D_\mu$, and
$\sigma^{\mu\nu}=i[\gamma^\mu,\gamma^\nu]/2$.

Motivated by the stringent experimental limits on
flavor-diagonal CP violation~\cite{Abel2020}, we restrict
our analysis to the CP-conserving sector with $\theta=0$.

\subsection{Covariant deformation}
\label{subsec:ordering-family}

The free correction $\Box\,\partial_\mu$ contains three commuting
derivatives. Replacing them by covariant derivatives introduces
an ordering ambiguity, since $[D_\mu,D_\nu]=-igF_{\mu\nu}$.
The three metric-contracted orderings share the same free limit
but differ by curvature-dependent terms. We therefore consider
\begin{equation}
\begin{aligned}
D'_\mu={}&D_\mu-\beta\bigl(
a\,D^2D_\mu+b\,D^\nu D_\mu D_\nu
+c\,D_\mu D^2\bigr)\\
&+\mathcal O(\beta^2),
\end{aligned}
\label{eq:three-ordering-family}
\end{equation}
with dimensionless coefficients $a,b,c$ and an implicit spinor
identity. Each ordering is gauge covariant, while matching to
Eq.~\eqref{eq:free-deformation} requires
\begin{equation}
a+b+c=1.
\label{eq:ordering-free-condition}
\end{equation}

Hermiticity constrains the allowed coefficients. Let $\dagger$
denote the formal adjoint under the spacetime integral and define
\begin{equation}
\mathcal O^\ddagger=\gamma^0\mathcal O^\dagger\gamma^0.
\label{eq:formal-dirac-adjoint}
\end{equation}
Neglecting surface terms,
\begin{equation}
D_\mu^\ddagger=-D_\mu,\qquad
(D^2)^\ddagger=D^2,\qquad
(i\slashed D)^\ddagger=i\slashed D.
\label{eq:formal-adjoint-rules}
\end{equation}
Since $(D^2D_\mu)^\ddagger=-D_\mu D^2$ and
$D^\nu D_\mu D_\nu$ is anti-Hermitian, Hermiticity of the
fermion action requires
\begin{equation}
c=a^*,\qquad b\in\mathbb R.
\label{eq:general-ordering-hermiticity}
\end{equation}
We restrict our analysis to real coefficients. Within
this family,
\begin{equation}
a=c=\frac{1-b}{2},
\label{eq:ordering-hermiticity}
\end{equation}
leaving $b$ unrestricted.

Using
\begin{equation}
\begin{aligned}
&D^2D_\mu+D_\mu D^2-2D^\nu D_\mu D_\nu\\
&\qquad=[D^\nu,[D_\nu,D_\mu]]=igJ_\mu,
\end{aligned}
\label{eq:ordering-commutator-identity}
\end{equation}
we can write
\begin{equation}
\begin{aligned}
D'_\mu&=D_\mu+\beta R_\mu+\mathcal O(\beta^2),\\
R_\mu&=-\frac12\{D^2,D_\mu\}+\frac{ig\,b}{2}J_\mu.
\end{aligned}
\label{eq:hermitian-ordering-deformation}
\end{equation}
Both terms in $R_\mu$ are anti-Hermitian, and the ordering
dependence is contained entirely in the term proportional
to $J_\mu$. The operator satisfies
\begin{equation}
\begin{aligned}
D'_\mu&\longrightarrow U D'_\mu U^{-1},\\
\left.D'_\mu\right|_{A=0}
&=(1-\beta\,\Box)\partial_\mu+\mathcal O(\beta^2).
\end{aligned}
\label{eq:free-matching-condition}
\end{equation}
We keep $b$ arbitrary and work at tree level through first
order in $\beta$.
Throughout, $\doteq$ denotes equality modulo total derivatives
and $\simeq$ equivalence under perturbative local field
redefinitions.

\section{Fermion sector}
\label{sec:fermion-sector}

Substituting Eq.~\eqref{eq:hermitian-ordering-deformation}
into the fermion action gives
\begin{equation}
\begin{aligned}
\mathcal L'_{\mathrm m}
={}&\bar\psi(i\slashed D-m)\psi\\
&-\frac{i\,\beta}{2}\bar\psi\{\slashed D,D^2\}\psi\\
&-\frac{\beta\, g\,b}{2}\bar\psi\gamma^\mu J_\mu\psi
+\mathcal O(\beta^2).
\end{aligned}
\label{eq:matter-lagrangian-expansion}
\end{equation}
To remove the higher derivatives, define
\begin{equation}
K\equiv i\slashed D,\qquad H_0\equiv K-m,\qquad
S\equiv\sigma^{\mu\nu}F_{\mu\nu}.
\label{eq:fermion-reduction-notation}
\end{equation}
The Clifford algebra and $[D_\mu,D_\nu]=-igF_{\mu\nu}$ imply
\begin{equation}
K^2=-D^2+\frac g2S.
\label{eq:dirac-operator-square}
\end{equation}
We perform the gauge-covariant field redefinition
\begin{equation}
\psi=(1+\beta\,\mathcal R)\,\chi,
\qquad
\mathcal R=\frac12D^2-\frac m2(K+m),
\label{eq:fermion-field-redefinition}
\end{equation}
with the corresponding transformation of $\bar\psi$.
Since $\mathcal R^\ddagger=\mathcal R$, the correction satisfies
\begin{equation}
-\frac12\{K,D^2\}
+\mathcal R^\ddagger H_0+H_0\mathcal R
=m^3-\frac{g\,m}{2}S.
\label{eq:exact-fermion-reduction}
\end{equation}
The higher-derivative term thus reduces to a mass shift and a
Pauli interaction, while the $b$-dependent term is unchanged
at this order. The derivation is given in Appendix~\ref{app:redefinitions}.

Defining
\begin{equation}
m_{\mathrm{eff}}\equiv m-\beta \,m^3,
\label{eq:effective-fermion-mass}
\end{equation}
we obtain
\begin{equation}
\begin{aligned}
\mathcal L'_{\mathrm m}\simeq{}&
\bar\chi(i\slashed D-m_{\mathrm{eff}})\chi\\
&-\frac{\beta \,g\, m_{\mathrm{eff}}}{2}
\bar\chi\sigma^{\mu\nu}F_{\mu\nu}\chi\\
&-\frac{\beta \,g\,b}{2}\bar\chi\gamma^\mu J_\mu\chi
+\mathcal O(\beta^2).
\end{aligned}
\label{eq:reduced-fermion-lagrangian}
\end{equation}
Replacing $m$ by $m_{\mathrm{eff}}$ in the dipole changes only
terms of order $\beta^2$. The mass shift is absorbed into the
physical input mass and does not by itself constrain $\beta$.
The dipole coefficient is independent of $b$ and vanishes in
the massless limit, as required by chiral symmetry.

\section{Gauge sector}
\label{sec:gauge-sector}

\subsection{Generalized curvature}
\label{subsec:generalized-curvature}

The commutator of the deformed derivatives defines
\begin{equation}
\begin{aligned}
\mathscr F_{\mu\nu}
&\equiv\frac{i}{g}[D'_\mu,D'_\nu]
=F_{\mu\nu}+\beta\mathscr K_{\mu\nu}
+\mathcal O(\beta^2),\\
\mathscr K_{\mu\nu}
&=\frac{i}{g}\bigl([D_\mu,R_\nu]-[D_\nu,R_\mu]\bigr),
\end{aligned}
\label{eq:generalized-curvature}
\end{equation}
which transforms covariantly,
$\mathscr F_{\mu\nu}\to U\mathscr F_{\mu\nu}U^{-1}$.
Using Eq.~\eqref{eq:hermitian-ordering-deformation}, we obtain
\begin{align}
\mathscr K_{\mu\nu}
={}&-\{D^2,F_{\mu\nu}\}
\nonumber\\
&-\frac12\bigl\{\{D^\rho,F_{\mu\rho}\},D_\nu\bigr\}
\nonumber\\
&+\frac12\bigl\{\{D^\rho,F_{\nu\rho}\},D_\mu\bigr\}
\nonumber\\
&-\frac b2(\nabla_\mu J_\nu-\nabla_\nu J_\mu).
\label{eq:reference-curvature-operator}
\end{align}
Unlike $F_{\mu\nu}$, the generalized curvature is a differential
operator, containing up to two covariant derivatives acting
to the right. Its internal trace therefore does not by itself
define a local gauge Lagrangian.

The Jacobi identity gives the operator Bianchi identity,
\begin{equation}
[D'_\rho,\mathscr F_{\mu\nu}]
+[D'_\mu,\mathscr F_{\nu\rho}]
+[D'_\nu,\mathscr F_{\rho\mu}]=0,
\label{eq:generalized-operator-bianchi}
\end{equation}
valid order by order in $\beta$.

\subsection{Local gauge action}
\label{subsec:gauge-prescription}

The generalized curvature contains derivatives acting to the
right, so its square is a differential operator rather than
a local Lagrangian. We evaluate this operator on a parallel
transporter, whose endpoint transformation law ensures that
the coincidence limit is gauge covariant. We choose the
straight transporter,
\begin{equation}
\mathcal W(x,y)=\mathcal P\exp\!\left[
ig\int_0^1ds\,(x-y)^\rho A_\rho(y+s(x-y))
\right],
\label{eq:straight-parallel-transporter}
\end{equation}
which transforms as
$\mathcal W(x,y)\to U(x)\mathcal W(x,y)U^{-1}(y)$.
Here $\mathcal P$ denotes path ordering, with larger values
of $s$ placed to the left.
For a covariant differential operator $\widehat O$, define
\begin{equation}
\mathscr E_{\mathcal W}[\widehat O](x)
=\left.\widehat O_x\mathcal W(x,y)\right|_{y=x}.
\label{eq:curvature-coincidence-map}
\end{equation}
Derivatives act on $x$ at fixed $y$ before taking coincidence.

We define the gauge Lagrangian by
\begin{equation}
\begin{aligned}
\mathcal L_{\mathrm g}^{\mathcal W}
&=-\frac12\mathcal I+\mathcal O(\beta^2),\\
\mathcal I
&=\operatorname{Re}\operatorname{tr}_g
\mathscr E_{\mathcal W}
[\mathscr F_{\mu\nu}\mathscr F^{\mu\nu}],
\end{aligned}
\label{eq:prescribed-gauge-lagrangian}
\end{equation}
with the curvature product expanded through first order in
$\beta$. The normalization reproduces the Yang--Mills
Lagrangian at $\beta=0$. For the Abelian theory, we omit
the internal trace and replace the factor $-1/2$ by $-1/4$.
A general proof of gauge invariance is given in
Appendix~\ref{app:kober-test}.

At coincident endpoints, $\mathcal W(x,x)=\mathbf1_g$, where
$\mathbf1_g$ is the identity in gauge space. Along the straight
segment, the transporter satisfies
\begin{equation}
(x-y)^\rho\partial_\rho^x\mathcal W(x,y)
=ig(x-y)^\rho A_\rho(x)\mathcal W(x,y).
\end{equation}
Hence $(x-y)^\rho D_\rho^x\mathcal W(x,y)=0$.
Differentiating this relation at fixed $y$ and using
$[D_\rho,D_\sigma]=-igF_{\rho\sigma}$ gives, at coincidence,
\begin{equation}
\begin{aligned}
\left.D^x_\rho\mathcal W\right|_{y=x}&=0,\\
\left.D^x_{(\rho}D^x_{\sigma)}\mathcal W\right|_{y=x}&=0,\\
\left.D^x_\rho D^x_\sigma\mathcal W\right|_{y=x}
&=-\frac{ig}{2}F_{\rho\sigma}.
\end{aligned}
\label{eq:transporter-coincidence-identities}
\end{equation}
These identities are standard ingredients of covariant
derivative expansions~\cite{Kuzenko2003,FuentesMartin2025Functional}. For a local
gauge-matrix coefficient $M$, the Leibniz rule gives
\begin{align}
\mathscr E_{\mathcal W}[M]&=M,
\nonumber\\
\mathscr E_{\mathcal W}[D_\rho\circ M]&=\nabla_\rho M,
\nonumber\\
\mathscr E_{\mathcal W}[D_\rho D_\sigma\circ M]
&=\nabla_\rho\nabla_\sigma M
-\frac{ig}{2}M F_{\rho\sigma}.
\label{eq:left-ordered-coincidence-rules}
\end{align}
These rules suffice to evaluate the action at order $\beta$.
Since $\mathscr E_{\mathcal W}$ does not preserve products,
we evaluate the complete curvature product before using
cyclicity of the local trace or integrating by parts.

\subsection{Gauge operators}
\label{subsec:gauge-matching}

To evaluate the gauge action, define
\begin{equation}
\begin{aligned}
\mathcal K&\equiv\operatorname{tr}_g
[(\nabla_\rho F_{\mu\nu})(\nabla^\rho F^{\mu\nu})],\\
\mathcal J&\equiv\operatorname{tr}_g(J_\mu J^\mu).
\end{aligned}
\label{eq:gauge-derivative-invariants}
\end{equation}
Applying Eq.~\eqref{eq:left-ordered-coincidence-rules} to
the curvature product gives
\begin{equation}
\begin{aligned}
\mathcal I\doteq{}&
\operatorname{tr}_g(F_{\mu\nu}F^{\mu\nu})+2\beta[\mathcal K+(1-b)\mathcal J]
+\mathcal O(\beta^2).
\end{aligned}
\label{eq:matched-gauge-invariant}
\end{equation}
The derivation, including both operator orderings
$F_{\mu\nu}\mathscr K^{\mu\nu}$ and
$\mathscr K_{\mu\nu}F^{\mu\nu}$, is given in
Appendix~\ref{app:curvature}.

The Bianchi identity relates $\mathcal K$ to $\mathcal J$
and the CP-even cubic operator~\cite{Grzadkowski2010Warsaw}
\begin{equation}
\mathcal O_{3F}
\equiv f^{ABC}F^A_\mu{}^\nu
F^B_\nu{}^\rho F^C_\rho{}^\mu.
\label{eq:cubic-gauge-operator-definition}
\end{equation}
Specifically,
\begin{align}
\nabla^2F_{\mu\nu}
={}&-\nabla_\mu J_\nu+\nabla_\nu J_\mu
\nonumber\\
&-2ig[F_{\mu\rho},F_\nu{}^\rho],
\nonumber\\
\mathcal K&\doteq2\mathcal J-g\mathcal O_{3F}.
\label{eq:gauge-bianchi-reduction}
\end{align}
The gauge Lagrangian therefore becomes
\begin{equation}
\begin{aligned}
\mathcal L_{\mathrm g}^{\mathcal W}\doteq{}&
-\frac14F^A_{\mu\nu}F^{A\mu\nu}\\
&+\beta\left[
-\frac{3-b}{2}J^A_\mu J^{A\mu}
+g\mathcal O_{3F}\right]
+\mathcal O(\beta^2).
\end{aligned}
\label{eq:matched-gauge-lagrangian}
\end{equation}
No gauge-field equation of motion has been used.
The ordering dependence resides entirely in $J^2$, whose
reduction requires the fermion source and is treated next.

\section{Reduced effective theory}
\label{sec:reduced-eft}

\subsection{Ordering independence}
\label{subsec:combined-reduction}

Combining the fermion and gauge sectors, the leading
Lagrangian is
\begin{equation}
\mathcal L_0
=-\frac14F^A_{\mu\nu}F^{A\mu\nu}
+\bar\chi(i\slashed D-m_{\mathrm{eff}})\chi.
\label{eq:combined-leading-lagrangian}
\end{equation}
With the fermion current
\begin{equation}
j^{A\mu}\equiv\bar\chi\gamma^\mu T^A\chi,
\label{eq:fermion-gauge-current}
\end{equation}
its gauge-field variation reads
\begin{equation}
\delta\mathcal L_0
\doteq(-J^{A\mu}+g \,j^{A\mu})\,\delta A_\mu^A.
\label{eq:full-leading-gauge-variation}
\end{equation}
The leading gauge equation is therefore $J^{A\mu}=g\, j^{A\mu}$.

The terms involving $J_\mu^A$ in the order-$\beta$ correction are
\begin{equation}
-\frac{\beta\,(3-b)}{2}J^2
-\frac{\beta\, b\,g}{2}j\cdot J,
\label{eq:unreduced-current-pair}
\end{equation}
where Lorentz and adjoint indices are contracted.
We remove them through the field redefinition
\begin{equation}
A_{\mu,\mathrm{old}}^A
=A_\mu^A-\frac{\beta}{2}
\bigl[(3-b)J_\mu^A+3\,g\, j_\mu^A\bigr].
\label{eq:combined-gauge-field-redefinition}
\end{equation}
The shift transforms in the adjoint representation and hence
preserves the gauge transformation law of $A_\mu$. It induces
\begin{equation}
\delta\mathcal L_0\doteq
\frac{\beta\,(3-b)}{2}J^2
+\frac{\beta \,b\,g}{2}j\cdot J
-\frac{3\beta g^2}{2}j^2,
\label{eq:ordering-cancellation}
\end{equation}
canceling Eq.~\eqref{eq:unreduced-current-pair} for arbitrary
$b$ and leaving
\begin{equation}
\mathcal L_{4\chi}
=-\frac{3\,\beta\, g^2}{2}j_\mu^A j^{A\mu}.
\label{eq:single-species-four-fermion-lagrangian}
\end{equation}
The dipole and cubic gauge terms are unchanged through
order $\beta$.

Defining
\begin{equation}
\mathcal O_M=\bar\chi\sigma^{\mu\nu}F_{\mu\nu}\chi,
\qquad
\mathcal O_{VV}=j_\mu^A j^{A\mu},
\label{eq:reduced-matter-operator-definitions}
\end{equation}
we obtain the common reduced Lagrangian
\begin{equation}
\begin{aligned}
\mathcal L_{\mathrm{red}}\simeq{}&
\mathcal L_0+\beta\Bigl[
-\frac{g m_{\mathrm{eff}}}{2}\mathcal O_M\\
&-\frac{3g^2}{2}\mathcal O_{VV}
+g\mathcal O_{3F}
\Bigr]+\mathcal O(\beta^2).
\end{aligned}
\label{eq:ordering-independent-cp-even-reduced-lagrangian}
\end{equation}
This is our main matching result: all real orderings satisfying
$a+b+c=1$ and $a=c$ yield the same effective interactions.
Their coefficients are correlated and controlled by $\beta$
within the chosen gauge-action prescription.

\subsection{Several fermion species}
\label{subsec:multispecies}

We extend the construction to several Dirac fermions with
a common $\beta$ and ordering parameter $b$. For generators
$T_f^A$ in the representation of each species, the total
gauge current is
\begin{equation}
j_{\mathrm{tot}}^{A\mu}
=\sum_f\bar\chi_f\gamma^\mu T_f^A\chi_f.
\label{eq:multispecies-gauge-current}
\end{equation}
The leading gauge equation becomes
$J^{A\mu}=g \,j_{\mathrm{tot}}^{A\mu}$.
Using this current in
Eq.~\eqref{eq:combined-gauge-field-redefinition} gives
\begin{equation}
\begin{aligned}
\Delta\mathcal L_{\mathrm{multi}}
={}&-\frac{\beta\, g}{2}\sum_f m_{f,\mathrm{eff}}\,
\bar\chi_f\sigma^{\mu\nu}F_{\mu\nu}^A T_f^A\chi_f\\
&-\frac{3\,\beta\, g^2}{2}
j_{\mathrm{tot},\mu}^A \,j_{\mathrm{tot}}^{A\mu}
+\beta \,g\,\mathcal O_{3F}
+\mathcal O(\beta^2).
\end{aligned}
\label{eq:multispecies-reduced-interactions}
\end{equation}
The current square contains both same-species and mixed-species
contacts, with a factor of two for each distinct pair.
The dipoles are flavor diagonal and proportional to the
fermion masses.

\subsection{Abelian limit}
\label{subsec:abelian-limit}

For an Abelian gauge field, we use
\begin{equation}
\begin{aligned}
D_{\mu f}&=\partial_\mu-i\,e\,Q_f\, A_\mu,\\
j_{\mathrm{em}}^\mu
&=\sum_f Q_f\,\bar\chi_f\gamma^\mu\chi_f,
\end{aligned}
\label{eq:abelian-limit-conventions}
\end{equation}
where $Q_f$ is the charge of species $f$ in units of $e$.
The cubic field-strength contraction vanishes,
\begin{equation}
F_\mu{}^\nu F_\nu{}^\rho F_\rho{}^\mu=0,
\label{eq:abelian-cubic-identity}
\end{equation}
while the Bianchi identity implies
\begin{equation}
(\partial_\rho F_{\mu\nu})(\partial^\rho F^{\mu\nu})
\doteq2\,J_\mu\, J^\mu.
\label{eq:abelian-derivative-identity}
\end{equation}
With canonical Maxwell normalization, the reduced Lagrangian is
\begin{equation}
\begin{aligned}
\mathcal L_{\mathrm{red}}^{U(1)}\simeq{}&
-\frac14F_{\mu\nu}F^{\mu\nu}\\
&+\sum_f\bar\chi_f
(i\slashed D_f-m_{f,\mathrm{eff}})\chi_f\\
&-\frac{\beta \,e}{2}\sum_f Q_f\, m_{f,\mathrm{eff}}
\bar\chi_f\sigma^{\mu\nu}F_{\mu\nu}\chi_f\\
&-\frac{3\,\beta\, e^2}{2}
j_{\mathrm{em},\mu}\,j_{\mathrm{em}}^\mu
+\mathcal O(\beta^2).
\end{aligned}
\label{eq:abelian-reduced-lagrangian}
\end{equation}
The surviving corrections are magnetic dipoles and
vector-current contacts; no independent pure-photon
dimension-six operator remains.

\section{Phenomenological constraints}
\label{sec:phenomenology}
\label{sec:phenomenological-setup}

We use the effective interactions derived above to interpret
existing results on lepton magnetic moments, LEP contact
interactions, and dijet production. The matching is performed
at tree level, including the correlations among the induced
operators.

For the electromagnetic observables, we use pole lepton masses
and $\alpha(0)=e^2/(4\pi)$, and assume that weak interactions
remain unchanged. We consider the QCD sector separately.

\subsection{Magnetic moments}
\label{subsec:em-observables}

For a fermion of physical mass $m_f$, the induced Pauli
interaction is
\begin{equation}
\Delta\mathcal L_{\mathrm P}
=-\frac{\beta\, e\,Q_f\,m_f}{2}
\bar\chi_f\sigma^{\mu\nu}\chi_f F_{\mu\nu}.
\label{eq:induced-pauli-interaction}
\end{equation}
We define the electromagnetic form factors through
\begin{equation}
\mathcal V_f^\mu=i\,e\,Q_f\,\left[
\gamma^\mu F_{1f}(q^2)
+\frac{i\,\sigma^{\mu\nu}q_\nu}{2\,m_f}F_{2f}(q^2)\right],
\label{eq:electromagnetic-form-factor-convention}
\end{equation}
where $q=p'-p$ is the momentum transfer, while $p$ and $p'$ are the incoming and outgoing fermion
momenta, respectively.
The form factors satisfy $F_{1f}(0)=1$ and $a_f=F_{2f}(0)$.
Matching the Pauli term gives
\begin{equation}
\delta a_f=-\,2\,\beta\, m_f^2.
\label{eq:induced-anomalous-moment}
\end{equation}
This result is verified directly in the original field variables,
including external-state normalization, in
Appendix~\ref{app:electromagnetic-vertex}.

For a common $\beta$, the lepton contributions satisfy
\begin{equation}
\frac{\delta a_e}{m_e^2}
=\frac{\delta a_\mu}{m_\mu^2}
=\frac{\delta a_\tau}{m_\tau^2}=-\,2\,\beta.
\label{eq:universal-anomalous-moment-relation}
\end{equation}
They are negative for $\beta>0$ and grow quadratically with
the lepton mass.

The 2025 Muon $g-2$ Theory Initiative update reports
$\Delta a_\mu\equiv a_\mu^{\mathrm{exp}}-a_\mu^{\mathrm{SM}}
=(38\pm63)\times10^{-11}$, including the final Fermilab
result~\cite{Aliberti2025WP}.
Treating this residual as Gaussian gives the central 95\%
interval $[-85.48,161.48]\times10^{-11}$.
Requiring the predicted negative contribution to lie within
this interval yields
\begin{equation}
\Lambda_\beta
\gtrsim 5.1\ {\rm TeV}.
\label{eq:muon-numerical-limit}
\end{equation}
This bound assumes the quoted Standard Model prediction and
tree-level matching; it is a translation of the central interval,
not a one-sided Bayesian limit.

An electron constraint would require an independent measurement
of $\alpha$, rather than its extraction from $a_e$.
We do not combine the lepton channels.

\begin{table*}[t]
\caption{Tree-level matching to published constraints.
The muon bound uses a central Gaussian 95\% interval,
while the LEP bound retains the statistical convention
of the original analysis. The dijet entries are equivalent
scales for the four-quark interaction, not exclusions
of the full correlated theory.}
\label{tab:published-matching}
\begin{ruledtabular}
\begin{tabular}{llll}
Observable & Matching relation & Scale [TeV] & Assumptions\\
\hline
Muon $g-2$~\cite{Aliberti2025WP}
& $\delta a_\mu=-\,2\,\beta\, m_\mu^2$
& $\Lambda_\beta
\gtrsim5.1$
& Adopted SM prediction; central 95\% interval\\
LEP $VV^-$~\cite{Schael2013LEP}
& $\Lambda_\beta=\sqrt{3\,\alpha(0)}\Lambda_{\rm CI}$
& $\Lambda_\beta
\gtrsim3.0$
& Published 95\% limit; unchanged weak interactions\\
ATLAS dijets~\cite{Alioli2017QCD}
& $Z_{\rm QCD}=3\,m_W^2\,\beta$
& $7.1$
& Four-quark fit including quadratic terms\\
CMS dijets~\cite{Alioli2017QCD}
& $Z_{\rm QCD}=3\,m_W^2\,\beta$
& $5.7$
& Four-quark fit including quadratic terms\\
\end{tabular}
\end{ruledtabular}
\end{table*}

\subsection{LEP contact interactions}
\label{subsec:em-scattering}

For distinct fermion species, the electromagnetic current
square gives
\begin{equation}
\Delta\mathcal L_{ef}
=-\,3\,\beta\, e^2\,Q_e\,Q_f\,
(\bar e\gamma_\mu e)(\bar f\gamma^\mu f).
\label{eq:em-mixed-contact}
\end{equation}
For charged leptons, $Q_eQ_f=1$, matching the negative
vector--vector interaction in the LEP convention,
\begin{equation}
\Delta\mathcal L_{\rm CI}
=\frac{4\,\pi}{(1+\delta_{ef})\,\Lambda_{\rm CI}^2}
\sum_{i,j=L,R}\eta_{ij}\,
(\bar e_i\gamma_\mu e_i)(\bar f_j\gamma^\mu f_j),
\label{eq:lep-contact-convention}
\end{equation}
with $\eta_{ij}=-1$ for all four chiral combinations.
Here $i,j$ label chirality and $\delta_{ef}=1$ for $f=e$
and zero otherwise. Matching coefficients
gives
\begin{equation}
\Lambda_\beta=\sqrt{3\,\alpha(0)}\,\Lambda_{\rm CI}.
\label{eq:lep-scale-matching}
\end{equation}

The lepton-universal limit $\Lambda_{\rm CI}^{-}>20.0$ TeV
in Table~3.15 of Ref.~\cite{Schael2013LEP} therefore implies
\begin{equation}
\Lambda_\beta
\gtrsim 3.0\ {\rm TeV}.
\label{eq:lep-bound}
\end{equation}
The bound retains the 95\% statistical convention of the LEP
analysis and assumes unchanged weak interactions. The dipole
terms give mass-suppressed corrections to these processes.

At $\sqrt{s}=209$ GeV, the bound implies $\beta\, s<5\cdot 10^{-3}$,
consistent with a perturbative expansion. The LEP fit also
includes terms quadratic in the contact interaction; we use
its published limit without repeating the fit.

\subsection{Dijet production}
\label{subsec:strong-observables}

For quarks in the fundamental representation, the effective
QCD interactions are
\begin{equation}
\begin{aligned}
\Delta\mathcal L_{\rm QCD}={}&
-\frac{\beta\,g_s}{2}\sum_q m_q\,
\bar q\sigma^{\mu\nu}T^Aq\,G^A_{\mu\nu}\\
&-\frac{3\,\beta\,g_s^2}{2}j_c^{A\mu}j^A_{c\mu}
+\beta g_s\mathcal O_{3G},
\end{aligned}
\label{eq:qcd-pheno-lagrangian}
\end{equation}
where $g_s$ is the strong coupling, $G^A_{\mu\nu}$ is the
gluon field strength, and
$j_c^{A\mu}=\sum_q\bar q\gamma^\mu T^Aq$.
The operator $\mathcal O_{3G}$ follows the definition in
Eq.~\eqref{eq:cubic-gauge-operator-definition}.

\begin{figure}[t]
    \centering
    \includegraphics[width=\columnwidth]{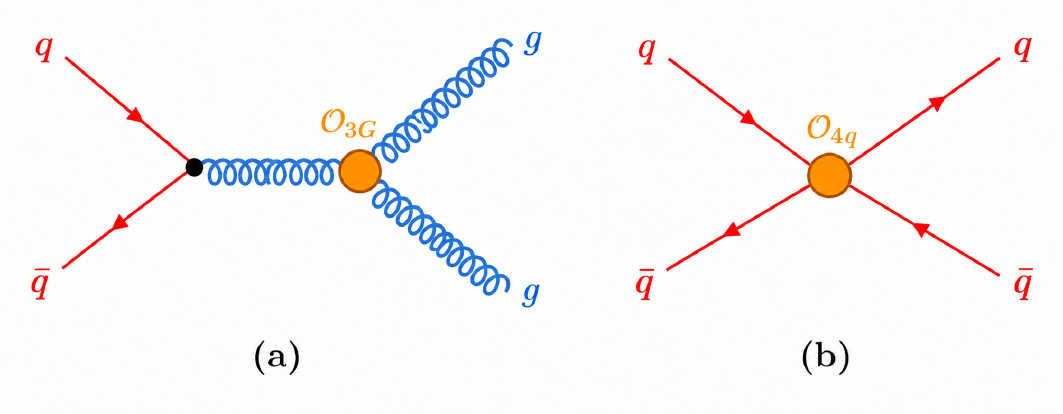}
\caption{
Representative contributions to dijet production:
(a) $q\bar q\to gg$ with an insertion of $\mathcal O_{3G}$;
(b) $q\bar q\to q\bar q$ through
$\mathcal O_{4q}\equiv j_c^{A\mu}j^A_{c\mu}$,
with coefficient $-\,3\,\beta \,g_s^2/2$.
Orange circles denote effective-operator insertions;
the black dot denotes the standard quark--gluon vertex.
}
    \label{fig:CPeven_probes}
\end{figure}

Representative contributions to dijet production are shown
in Fig.~\ref{fig:CPeven_probes}.

The dijet analysis of Ref.~\cite{Alioli2017QCD} constrains
the operator
\begin{equation}
\Delta\mathcal L_Z
=-\frac{Z_{\rm QCD}}{2\,m_W^2}
(\nabla_\mu G^{A\mu\nu})^2
\simeq-\frac{Z_{\rm QCD}g_s^2}{2\,m_W^2}j_c^2.
\label{eq:qcd-oblique-convention}
\end{equation}
Matching its four-quark coefficient gives
\begin{equation}
Z_{\rm QCD}=3\,m_W^2\,\beta,
\qquad
\Lambda_\beta=m_W\,\sqrt{\frac{3}{Z_{\rm QCD}}}.
\label{eq:qcd-Z-matching}
\end{equation}
The $W$-boson mass enters only as a normalization scale.

For massless quarks, the dipole terms vanish and
$\mathcal O_{3G}$ does not interfere with the QCD amplitude
in tree-level $2\to2$ scattering~\cite{DixonShadmi1994}.
The four-quark interaction therefore gives the leading
correction to dijet production at first order in $\beta$
in this limit. Beyond tree level, cubic-gluon interference
can also contribute~\cite{Hirschi2018Gluon}.

For the ATLAS and CMS 7 TeV analyses, Table~3 of
Ref.~\cite{Alioli2017QCD} gives upper limits on positive
$Z_{\rm QCD}$ of $3.9\times10^{-4}$ and $6.0\times10^{-4}$,
respectively. These results use the first rapidity bin and
parton distributions fitted without jet data.
With $m_W=80.4$ GeV, Eq.~\eqref{eq:qcd-Z-matching}
converts them to
\begin{equation}
\begin{aligned}
\Lambda_\beta^{\rm ATLAS}&=7.1\ {\rm TeV},\\
\Lambda_\beta^{\rm CMS}&=5.7\ {\rm TeV}.
\end{aligned}
\label{eq:qcd-template-scales}
\end{equation}

These values indicate the sensitivity of the published fits
to our four-quark interaction, rather than bounds on the
complete theory. The fits include terms quadratic in
$Z_{\rm QCD}$, whereas our matching retains only first order
in $\beta$. At quadratic order, the other induced interactions
and interference with dimension-eight operators must also
be considered. A bound on the correlated theory therefore
requires a dedicated fit.

Table~\ref{tab:published-matching} summarizes the results.
We quote each constraint separately, without a combined fit.

\section{Discussion and conclusions}
\label{sec:conclusions}

We have constructed a local gauge-invariant effective theory
for a GUP-inspired deformation using a straight parallel
transporter. For the real three-ordering family satisfying
$a+b+c=1$ and $a=c$, the remaining ordering parameter is
removed by a field redefinition at first order in $\beta$.
The combined reduction of the matter and gauge sectors yields
correlated magnetic dipoles, four-fermion vector-current
operators, and a CP-even triple-gauge interaction.
The dipoles give $\delta a_f=-\,2\,\beta\, m_f^2$, independently
of the ordering.

The construction addresses the distinction between a
gauge-covariant differential operator and a gauge-invariant
local action. We have shown that the Abelian action displayed
in Eq.~(22) of Ref.~\cite{Kober:2010sj}, interpreted as written,
fails the off-shell gauge identity. The transporter prescription
provides a local action invariant under general Abelian and
non-Abelian gauge transformations. It defines an interacting
realization, rather than a unique consequence of the free GUP.

At tree level, the muon magnetic moment gives
$\Lambda_\beta\geq5.1$ TeV using the adopted Standard Model
prediction and a central Gaussian 95\% interval.
The published 95\% LEP contact limit gives
$\Lambda_\beta>3$ TeV with unchanged weak interactions.
Matching the four-quark interaction to ATLAS and CMS dijet
fits yields equivalent scales of $7.1$ and $5.7$ TeV,
respectively. These indicate collider sensitivity but are
not bounds on the full correlated theory, which requires
a dedicated analysis.

An electroweak completion would extend the matching
to SMEFT and allow the correlations to be tested
with electroweak and flavor observables.
A one-loop study of a related massless GUP-deformed
QED theory has found a radiatively generated
axial--axial four-fermion
interaction~\cite{Valero2026QEDRen}.
Extending the renormalization analysis to the present
transporter-defined action, including massive
fermions and the non-Abelian sector, would test
the radiative stability of the correlated
coefficients and determine their scale evolution.

\appendix

\section{Field redefinitions}
\label{app:redefinitions}

For the fermion redefinition in
Eq.~\eqref{eq:fermion-field-redefinition}, direct expansion gives
\begin{equation}
\begin{aligned}
\mathcal R^\ddagger \,H_0+H_0\,\mathcal R
={}&\frac12\{D^2,K\}-m\,D^2\\
&-m\,K^2+m^3.
\end{aligned}
\label{eq:appendix-fermion-shift}
\end{equation}
The anticommutator cancels the higher-derivative term in
Eq.~\eqref{eq:matter-lagrangian-expansion}. Using
$K^2=-D^2+g\,S/2$, the remaining terms give
\begin{equation}
-m\,(D^2+K^2)+m^3
=m^3-\frac{g\,m}{2}S,
\label{eq:appendix-fermion-reduction}
\end{equation}
establishing Eq.~\eqref{eq:exact-fermion-reduction}
without using equations of motion. The $b$-dependent
interaction changes only at order $\beta^2$.

The gauge-field redefinition in
Eq.~\eqref{eq:combined-gauge-field-redefinition} induces
\begin{equation}
\begin{aligned}
\delta\mathcal L_0
&\doteq-\frac{\beta}{2}(-J+gj)\cdot[(3-b)J+3\,g\,j]\\
&=\frac{\beta(3-b)}2J^2
+\frac{\beta\, b\,g}2j\cdot J-\frac{3\,\beta\, g^2}2j^2.
\end{aligned}
\label{eq:appendix-gauge-shift}
\end{equation}
This cancels the $J^2$ and $j\cdot J$ terms, leaving
a contact interaction independent of $b$. For several
fermion species, $j$ is replaced by the total current.

\section{Gauge-sector matching}
\label{app:curvature}

We decompose the curvature correction as
$\mathscr K_{\mu\nu}=\mathscr K^{(1)}_{\mu\nu}
+\mathscr K^{(2)}_{\mu\nu}+\mathscr K^{(J)}_{\mu\nu}$, with
\begin{align}
\mathscr K^{(1)}_{\mu\nu}
&=-\{D^2,F_{\mu\nu}\},
\nonumber\\
\mathscr K^{(2)}_{\mu\nu}
&=-\frac12\bigl\{\{D^\rho,F_{\mu\rho}\},D_\nu\bigr\}
\nonumber\\
&\quad+\frac12\bigl\{\{D^\rho,F_{\nu\rho}\},D_\mu\bigr\},
\nonumber\\
\mathscr K^{(J)}_{\mu\nu}
&=-\frac b2(\nabla_\mu J_\nu-\nabla_\nu J_\mu).
\label{eq:appendix-curvature-split}
\end{align}
The order-$\beta$ curvature product contains both
$F^{\mu\nu}\mathscr K_{\mu\nu}$ and
$\mathscr K_{\mu\nu}F^{\mu\nu}$.

For the first contribution, the coincidence identities give
\begin{equation}
\begin{aligned}
&\operatorname{Re}\operatorname{tr}_g
\mathscr E_{\mathcal W}
[F^{\mu\nu}\mathscr K^{(1)}_{\mu\nu}
+\mathscr K^{(1)}_{\mu\nu}F^{\mu\nu}]\\
&\quad\doteq
-2\operatorname{tr}_g(F^{\mu\nu}\nabla^2F_{\mu\nu})
\doteq2\,\mathcal K.
\end{aligned}
\label{eq:appendix-curvature-a-matching}
\end{equation}
Here the trace of $\nabla^2(F^{\mu\nu}F_{\mu\nu})$
has been discarded as a total derivative.

For local gauge-matrix coefficients $M,N$, we use
\begin{equation}
\begin{aligned}
&\operatorname{tr}_g\mathscr E_{\mathcal W}
\Big[N\{\{D_\alpha,M\},D_\lambda\}
+\{\{D_\alpha,M\},D_\lambda\}N\Big]\\
&\quad\doteq
-2\operatorname{tr}_g
[(\nabla_\lambda N)(\nabla_\alpha M)].
\end{aligned}
\label{eq:appendix-nested-anticommutator}
\end{equation}
This follows by moving derivatives to the right, applying
Eq.~\eqref{eq:transporter-coincidence-identities}, and
integrating by parts after evaluating the local trace.
Applying it to $\mathscr K^{(2)}$ yields
\begin{equation}
\begin{aligned}
&\operatorname{Re}\operatorname{tr}_g
\mathscr E_{\mathcal W}
[F^{\mu\nu}\mathscr K^{(2)}_{\mu\nu}
+\mathscr K^{(2)}_{\mu\nu}F^{\mu\nu}]\\
&\quad\doteq
2\operatorname{tr}_g
[(\nabla_\nu F^{\mu\nu})(\nabla^\rho F_{\mu\rho})]
=2\,\mathcal J.
\end{aligned}
\label{eq:appendix-curvature-b-matching}
\end{equation}
The remaining contribution is
\begin{equation}
\begin{aligned}
&\operatorname{Re}\operatorname{tr}_g
\mathscr E_{\mathcal W}
[F^{\mu\nu}\mathscr K^{(J)}_{\mu\nu}
+\mathscr K^{(J)}_{\mu\nu}F^{\mu\nu}]\\
&\quad=-2b\operatorname{tr}_g(F^{\mu\nu}\nabla_\mu J_\nu)
\doteq-2\,b\,\mathcal J,
\end{aligned}
\label{eq:appendix-curvature-J-matching}
\end{equation}
where $\nabla_\mu F^{\mu\nu}=-J^\nu$.
The sum reproduces Eq.~\eqref{eq:matched-gauge-invariant}
without using gauge-field equations of motion.

Finally, integration by parts and the Bianchi identity give
\begin{equation}
\begin{aligned}
\mathcal K
&\doteq-\operatorname{tr}_g(F^{\mu\nu}\nabla^2F_{\mu\nu})\\
&\doteq2\,\mathcal J
+2\,i\,g\,\operatorname{tr}_g
\bigl(F^{\mu\nu}[F_{\mu\rho},F_\nu{}^\rho]\bigr).
\end{aligned}
\label{eq:appendix-gauge-bianchi-trace}
\end{equation}
Using $[T^A,T^B]=if^{ABC}T^C$ and
$\operatorname{tr}_g(T^AT^B)=\delta^{AB}/2$, the last term
is $-\,g\,\mathcal O_{3F}$, giving
$\mathcal K\doteq2\,\mathcal J-g\,\mathcal O_{3F}$
with the normalization of
Eq.~\eqref{eq:cubic-gauge-operator-definition}.

\section{Electromagnetic vertex}
\label{app:electromagnetic-vertex}

We verify the dipole matching directly in the original field
variables. For one Abelian fermion, the inverse free propagator,
with the overall Feynman factor suppressed, is
\begin{equation}
S_0^{-1}(p)=(1+\beta p^2)\,\slashed p-m+\mathcal O(\beta^2).
\label{eq:unreduced-fermion-inverse-propagator}
\end{equation}
Expanding around the pole connected to the ordinary Dirac
solution gives
\begin{equation}
\begin{aligned}
m&=m_{\mathrm p}+\beta \,m_{\mathrm p}^3
+\mathcal O(\beta^2),\\
Z_{\mathrm p}&=1-3\,\beta\, m_{\mathrm p}^2
+\mathcal O(\beta^2),
\end{aligned}
\label{eq:unreduced-mass-and-residue}
\end{equation}
where $m_{\mathrm p}$ is the physical mass and $Z_{\mathrm p}$
is the residue multiplying
$i(\slashed p+m_{\mathrm p})/(p^2-m_{\mathrm p}^2)$.

Let $p$ and $p'$ denote the incoming and outgoing fermion
momenta, with $q=p'-p$. Expanding the matter action to first
order in the photon field gives the vertex
$ieQ\Gamma^\mu_{\mathrm{bare}}$, where
\begin{equation}
\begin{aligned}
\Gamma^\mu_{\mathrm{bare}}(p',p)
={}&\gamma^\mu+\frac{\beta}{2}
\Bigl[(p'^2+p^2)\gamma^\mu\\
&\qquad+(p'+p)^\mu(\slashed p'+\slashed p)\Bigr]\\
&-\frac{\beta\, b}{2}(q^2\gamma^\mu-q^\mu\slashed q)
+\mathcal O(\beta^2).
\end{aligned}
\label{eq:unreduced-electromagnetic-vertex}
\end{equation}
The $b$-dependent term is transverse, and the vertex satisfies
the off-shell Ward identity
\begin{equation}
q_\mu\Gamma^\mu_{\mathrm{bare}}
=S_0^{-1}(p')-S_0^{-1}(p)
\label{eq:unreduced-ward-identity}
\end{equation}
through order $\beta$.

For on-shell spinors of mass $m_{\mathrm p}$, the Gordon
identity reads
\begin{equation}
\begin{aligned}
\bar u(p')(p'+p)^\mu u(p)
=\bar u(p')\Bigl[
2\,m_{\mathrm p}\gamma^\mu-i\,\sigma^{\mu\nu}q_\nu
\Bigr]u(p).
\end{aligned}
\label{eq:gordon-identity-convention}
\end{equation}
Using this identity and $\bar u(p')\slashed q u(p)=0$, we obtain
\begin{equation}
\begin{aligned}
&\bar u(p')\Gamma^\mu_{\mathrm{bare}}u(p)\\
&=\bar u(p')\Bigl[
(1+3\,\beta\, m_{\mathrm p}^2-\beta\, b\,q^2/2)\gamma^\mu\\
&\hspace{20mm}-i\,\beta\, m_{\mathrm p}\,\sigma^{\mu\nu}q_\nu
\Bigr]u(p)+\mathcal O(\beta^2).
\end{aligned}
\label{eq:unreduced-on-shell-vertex}
\end{equation}
Each external leg contributes $Z_{\mathrm p}^{1/2}$.
Multiplication by $Z_{\mathrm p}$ cancels the
$3\beta m_{\mathrm p}^2$ term, while the ordering-dependent
contribution vanishes at $q^2=0$. The normalized form factors
therefore satisfy
\begin{equation}
F_1(0)=1,\qquad
\delta F_2(0)=-2\,\beta\, m_{\mathrm p}^2.
\label{eq:unreduced-normalized-form-factors}
\end{equation}
With $m_{\mathrm p}=m_{\mathrm{eff}}+\mathcal O(\beta^2)$,
this reproduces Eq.~\eqref{eq:induced-anomalous-moment}
for arbitrary $b$.

\section{Gauge invariance}
\label{app:kober-test}

\subsection{Previously proposed Abelian action}

We examine the local Abelian action in Eq.~(22) of
Ref.~\cite{Kober:2010sj}, using its real
gauge potential with the coupling absorbed into $A_\mu$.
No gauge condition or field equation is imposed.

For real fields, gauge invariance of the action requires
invariance of its real part. Define
\begin{equation}
\begin{aligned}
F^2&\equiv F_{\mu\nu}F^{\mu\nu},\\
Q_{\rho\sigma}
&\equiv\eta_{\rho\sigma}F^2
+2\,F_{\mu\rho}F^\mu{}_\sigma.
\end{aligned}
\label{eq:kober-general-tensors}
\end{equation}
The tensor $Q_{\rho\sigma}$ is symmetric and gauge invariant.
With the normalization of that reference,
\begin{equation}
\begin{aligned}
\operatorname{Re}\mathcal L_{\rm K}
={}&\frac14\,F^2-\beta F^{\mu\nu}\,\Box F_{\mu\nu}\\
&+\beta \,A^\rho\, Q_{\rho\sigma}A^\sigma
+\mathcal O(\beta^2).
\end{aligned}
\label{eq:kober-general-lagrangian}
\end{equation}
Only the last displayed term varies under
\begin{equation}
A_\mu\longrightarrow A_\mu+\partial_\mu\lambda,
\qquad F_{\mu\nu}\longrightarrow F_{\mu\nu}.
\label{eq:kober-general-transformation}
\end{equation}
For an infinitesimal parameter of compact support,
integration by parts gives
\begin{equation}
\delta_\lambda\operatorname{Re}S_{\rm K}
=-2\,\beta\int d^4x\,\lambda\,
\partial^\sigma(A^\rho Q_{\rho\sigma})
+\mathcal O(\beta^2).
\label{eq:kober-general-infinitesimal-variation}
\end{equation}
Gauge invariance would therefore require the off-shell identity
$\partial^\sigma(A^\rho Q_{\rho\sigma})\equiv0$.

Under a second gauge variation with parameter $\eta$,
this identity would imply
\begin{equation}
\begin{aligned}
0&\equiv\partial^\sigma
(Q_{\rho\sigma}\partial^\rho\eta)\\
&=Q_{\rho\sigma}\,\partial^\sigma\partial^\rho\eta
+(\partial^\sigma Q_{\rho\sigma})\,\partial^\rho\eta
\end{aligned}
\label{eq:kober-general-noether-obstruction}
\end{equation}
for every smooth $\eta$ of compact support. At an interior
point, its gradient can vanish while its symmetric Hessian
is arbitrary. This would require $Q_{\rho\sigma}\equiv0$.
However,
\begin{equation}
\eta^{\rho\sigma}Q_{\rho\sigma}=6\,F^2
\label{eq:kober-general-trace-obstruction}
\end{equation}
is not identically zero.

The displayed local action thus fails the off-shell gauge
identity, even after taking its real part. The obstruction
cannot be removed by a total derivative. This conclusion
concerns the local expression in Eq.~(22); the underlying
differential operator remains gauge covariant.

\subsection{Transporter construction}

For an arbitrary local gauge transformation $U(x)$,
\begin{equation}
\begin{aligned}
D'_\mu[A^U]&=U D'_\mu[A]U^{-1},\\
\mathscr F_{\mu\nu}[A^U]
&=U\mathscr F_{\mu\nu}[A]U^{-1}.
\end{aligned}
\end{equation}
These relations follow from covariance of each derivative
in the ordered products and hold for any ordering.
The transporter satisfies
\begin{equation}
\mathcal W[A^U](x,y)
=U(x)\mathcal W[A](x,y)U^{-1}(y).
\end{equation}
For a covariant differential operator $\widehat O$,
composition at fixed $y$ gives
\begin{equation}
\begin{aligned}
&\widehat O[A^U]_x\mathcal W[A^U](x,y)\\
&\qquad=U(x)
\bigl[\widehat O[A]_x\mathcal W[A](x,y)\bigr]U^{-1}(y),
\end{aligned}
\end{equation}
since derivatives with respect to $x$ do not act on
$U^{-1}(y)$. Taking coincidence after differentiation yields
\begin{equation}
\begin{aligned}
&\mathscr E_{\mathcal W[A^U]}[\widehat O[A^U]]=U(x)
\mathscr E_{\mathcal W[A]}[\widehat O[A]]U^{-1}(x).
\end{aligned}
\label{eq:general-coincidence-covariance}
\end{equation}
For $\widehat O=\mathscr F_{\mu\nu}\,\mathscr F^{\mu\nu}$,
the internal trace is invariant, and hence
\begin{equation}
\mathcal L_{\mathrm g}^{\mathcal W}[A^U]
=\mathcal L_{\mathrm g}^{\mathcal W}[A]
+\mathcal O(\beta^2).
\label{eq:general-prescribed-action-invariance}
\end{equation}
The matter action is also invariant under
$\psi\to U\,\psi$ and $\bar\psi\to\bar\psi\, U^{-1}$.
The full action is therefore gauge invariant off shell,
for both Abelian and non-Abelian groups.


\end{document}